# The Double Helix Nebula:
# a magnetic torsional wave propagating out of the Galactic centre

Mark Morris[1], Keven Uchida[2], and Tuan Do[1]

[1]*Department of Physics and Astronomy, University of California, Los Angeles, Los Angeles, CA 90095-1547, USA*
[2]*Center for Radiophysics and Space Research, Cornell University, Space Sciences Building, Ithaca, NY 14853-6801*

**Radioastronomical studies have indicated that the magnetic field in the central few hundred parsecs of our Milky Way Galaxy has a dipolar geometry and a strength substantially larger than elsewhere in the Galaxy, with estimates ranging up to a milligauss[1-6]. A strong, large-scale magnetic field can affect the Galactic orbits of molecular clouds by exerting a drag on them, it can inhibit star formation, and it can guide a wind of cosmic rays away from the central region, so a characterization of the magnetic field at the Galactic center is important for understanding much of the activity there. Here, we report Spitzer Space Telescope observations of an unprecedented infrared nebula having the morphology of an intertwined double helix. This feature is located about 100 pc from the Galaxy's dynamical centre toward positive Galactic latitude, and its axis is oriented perpendicular to the Galactic plane. The observed segment is about 25 pc in length, and contains about 1.25 full turns of each of the two continuous, helically wound strands. We interpret this feature as a torsional Alfvén wave propagating vertically away from the Galactic disk, driven by rotation of the magnetized circumnuclear gas disk. As such, it offers a new morphological probe of the Galactic center magnetic field. The direct connection between the circumnuclear disk and the double helix is**

**ambiguous, but the MSX images show a possible meandering channel that warrants further investigation.**

The new image was obtained using the MIPS camera[7] on the Spitzer Space Telescope at a wavelength of 24 μm. It reveals the elongated, double helix nebula (hereafter, DHN) shown in Figure 1. This nebula clearly extends beyond the edge of the observed field. The full extent of the DHN is evident in mid-infrared images previously obtained with the MSX satellite (ref. 9; their figure 10). In the MSX data (Figure 2), it is possible with hindsight to recognize the double helix, although the sensitivity and spatial resolution are not sufficient to have revealed this structure prior to the Spitzer observation. The MSX images show that the structure extends towards smaller Galactic latitude from the region observed with Spitzer; it is at least 20 arcminutes (50 parsecs) in length, extending between galactic coordinates $l = 0.08$, $b = 0.5$ and $l = 0.02$, $b = 0.80$, and it is possibly part of a larger structure (see below). However, there is no evidence in the MSX data that the helical strands persist outside the region observed with MIPS. It is also noteworthy that there is no sign in MSX images of a negative-latitude counterpart to the double helix nebula.

The four wavelengths observed with the MSX satellite (8.3, 12.1, 14.7, and 21.3 μm) provide a sampling of the spectral energy distribution of the mid-infrared emission. The surface brightnesses at a few sample locations, measured relative to the local background, uniformly indicate that the emission is substantially strongest at the shorter wavelengths (detailed in the online supplementary information). The best-fit color temperature is on the order of $630 \pm 40$K. If the emission corresponds to thermal emission from dust for which the mid-infrared emissivity is proportional to $\nu^p$, where $\nu$ is the frequency, then the implied best-fit dust temperature is $410 \pm 16$ K ($p$=1) or $310 \pm 10$ K ($p$=2). In either case, such high dust temperatures are normally found only near powerful heating sources such as supergiant stars or star clusters; they have not been seen over regions as extended as 50 pc, especially as high above the Galactic plane as the DHN, where there is no evidence for star formation. Two possible dust heating mechanisms which might be considered are: 1) heating by streaming of gas particles with respect to the dust (*e.g.*, as in a Galactic wind[10]) and 2) impulsive heating of small grains by ambient Galactic starlight. For reasonable densities of the streaming gas, the former mechanism provides inadequate heating, even for a relative gas-dust drift velocity as large as the Alfvén velocity in this region ($\sim 10^3$ km s$^{-1}$, see below). The second mechanism warrants further investigation. Ignoring dust extinction, the equilibrium temperature of dust in the Galactic bulge, 100 pc from the Galactic center, is only 30-40 K, but small grains can undergo large temperature excursions as they absorb UV photons[11], and the total mass required in small grains to account for the DHN emission, while sensitive to the assumed impingent UV intensity and grain size distribution, is much less than a solar mass.

Two potential alternatives to thermal dust emission -- thermal bremsstrahlung emission by hot electrons and nonthermal emission from a population of relativistic electrons – are both rendered unlikely by the observed spectral slope, as detailed in the supplementary on-line material. In any case, the average variance of the data from the best-fitting nonthermal power-law models is twice as large as that of the best-fitting thermal model: that with $p$=2. Consequently, we conclude that the emission observed from the double helix most likely arises from thermal dust emission. Shorter-wavelength observations in the 4 bands of the IRAC camera on Spitzer will help clarify this point.

The sky location and orientation of the DHN are very suggestive of an association with the center of the Galaxy. Not only does the long axis of this structure point roughly to the Galactic center, less than a degree away, but it is also oriented along the Galaxy's axis of rotation. Furthermore, as we argue below, an association with the Galactic center provides a natural explanation for this structure. At an assumed Galactic center distance of 8 kiloparsecs, the wavelength of the individual strands of the double helix – about 7.5 arcminutes – corresponds to a length of 19 pc, and the maximum strand separation, i.e., the overall width of the structure, is 1.4 arcminutes, or ~3.5 pc. The individual strands are just barely resolved, with a width of about 7 or 8 arcseconds at their narrowest locations, compared to the full width at half maximum of the MIPS point spread function at 24 μm: 6 arcseconds.

The observed helical structure is far too large to be attributed to stellar activity. Rather, it likely results from a dynamically ordered, large-scale, interstellar phenomenon involving interstellar gas, dust, and magnetic fields. We propose that the DHN is a magnetohydrodynamic torsional Alfvén wave propagating more or less vertically out of the Galactic plane, along magnetic field lines, from the near vicinity of the Galactic center. This hypothesis conforms to the apparent, global dipolar geometry of the Galactic center magnetic field[1]. It is natural to ascribe the driving of the torsional wave to rotation about the Galactic center, and an obvious candidate to do this is the circumnuclear disk (CND)[1,12-15]. The characteristics of the CND match those needed to produce the characteristics of the proposed torsional wave: it has an inner radius of ~1 pc, and extends out to several parsecs, being somewhat asymmetric in its outer regions, possibly because of an interaction with the SgrA East supernova remnant[1,16-18]. The lateral extent of the proposed torsional wave, ~3.5 pc, is consistent with the planar extent of the CND, 2 – 7 pc, and thus with the hypothesis that the rotation of the CND is responsible for a torsional wave propagating through a uniform field. Little lateral growth of the helical structure is expected as long as the Alfvén speed is much greater than the rotation speed of the driving disk, or as long as the longitudinal wavelength is much greater than the lateral extent of the feature, which is clearly the case. Furthermore, the CND is strongly magnetized, and its predominant shear-induced azimuthal field is believed to merge smoothly with the ambient vertical Galactic center field[19-21].

The rotation velocity of the CND is approximately constant at 100 km/s, giving a period of $10^4 R$(pc) years at radius $R$. Drawing a correspondence between the period of the CND and the wavelength of the double helix, we can derive the Alfvén velocity: $V_A = 10^3$ km/s. This, in turn, can be used to estimate the magnetic field strength, $B$, in this region, with $V_A^2 = B^2/(4\pi m_p n_p)$ where $n_p$ is the proton density in the medium through which the wave propagates, and $m_p$ is the proton mass. This gives $B = 0.5 n_p^{1/2}$ milligauss. The density in this region is not known; if the medium through which the wave propagates is the hot ($10^8$ K) medium evidenced in diffuse X-rays[22], then $n_p = \sim 0.1$ cm$^{-3}$, and consequently, $B = 0.1$ mG. The DHN is located on the outskirts of the region to which the $10^8$ K gas may extend[23], but given previous estimates for $B$ ranging from 0.01 to 1 mG[1,4,5], this is a plausible field strength. If, on the other hand, the field strength is that estimated from the rigidity of the nonthermal radio filaments, $B \sim 1$ mG[1], we infer a local density of $n_p \sim 5$ cm$^{-3}$, which does not violate any observational constraints.

A possible weakness of this hypothesis is that the torsional wave cannot yet be followed all the way down to its hypothetical source, the CND, presumably because of the enhanced confusion by intervening material and superimposed emission structures closer to the Galactic plane. The MSX data do, however, show a potential meandering channel along which the wave might propagate (Fig. 2). Such a meander itself can be ascribed to the kink instability, arising naturally in a twisted magnetic field. While the contrast of any such feature with the local background emission is weak, there are two locations at which evidence for a channel exists in the form of parallel, linear emission features having a separation (~5 pc) comparable to the width of the helical structure (marked with arrows in Fig. 2). These features (shown in greater detail in the supplementary online information) can be interpreted as limb-brightened cylinders having relatively thin, emitting walls. Patchy absorption abutting the brighter, lower-latitude one of these indicates the presence of a local concentration of dust and gas, so the emission from this apparent part of the channel might be attributable to an interaction between the magnetic energy in the torsional wave channel and the surrounding, relatively dense, interstellar material. One might also speculate that the meander of the apparent channel is partially caused by this interaction, deflecting the wave energy toward positive Galactic longitude. The absence of a negative-latitude counterpart is another potential weakness of the torsional wave hypothesis, inasmuch as such waves should propagate equally in both directions away from the driving disk, if that disk is symmetric about its midplane. However, the MSX images show a brighter more complex background at negative latitudes, so some combination of background confusion and dissipative shock interactions – both common in the Galactic center – could account for the absence of a counterpart.

One question that our hypothesis leaves unanswered is why the helical structure has two strands. A uniform, axisymmetric, rotating disk driving a torsional wave in a field perpendicular to the disk would produce a cylindrically symmetric structure. The presence of two strands indicates that the driver has an $m$=2 symmetry (surface density has a term of the form $e^{im\phi}$, where $\phi$ is the azimuthal angle). This could take the form of a bar, or, in the extreme case, one could attribute the strands of the double helix to two diametrically opposed blobs into which the vertical magnetic flux threading the disk had been concentrated. Since the time scale for propagation of an Alfvén wave from the CND to the observed double helix, ~100 pc away, is $10^5$ years, or $10/R$(pc) rotation periods of the disk, there has been sufficient time for the strong shear in the CND to have eliminated any $m$=2 deviation from axisymmetry that may have been present when the double helix was launched. Nonetheless, the CND is somewhat non-axisymmetric at the present time, and its inner portions have two prominent concentrations of material on opposite sides of the center along the major axis of the projected disk[24,25,12,14,15]. Consequently, it is possible that the $m$=2 deviation from axisymmetry is still in place, or that it has reformed since the double helix was launched.

Finally, if the favored mechanism for the mid-infrared emission from the DHN remains thermal dust emission, then we must face the question of why dust is present at all in the torsional wave. If the emission is from small dust grains, then such grains are likely to carry a net charge, in which case they can then be carried aloft by the torsional wave.

**Acknowledgements**   MM is grateful to Steven Cowley for illuminating discussions. This work is based on observations made with the Spitzer Space Telescope, which is operated by the Jet Propulsion Laboratory, California Institute of Technology under a contract with NASA.  Partial support for this work was provided by NASA through an award issued by JPL/Caltech. This research also made use of data products from the Midcourse Space Experiment, available via the NASA/ IPAC Infrared Science Archive.

**Author Information**  The authors declare that they have no competing financial interests.  Correspondence and requests for materials should be addressed to M.M. ([morris@astro.ucla.edu](mailto:morris@astro.ucla.edu)).

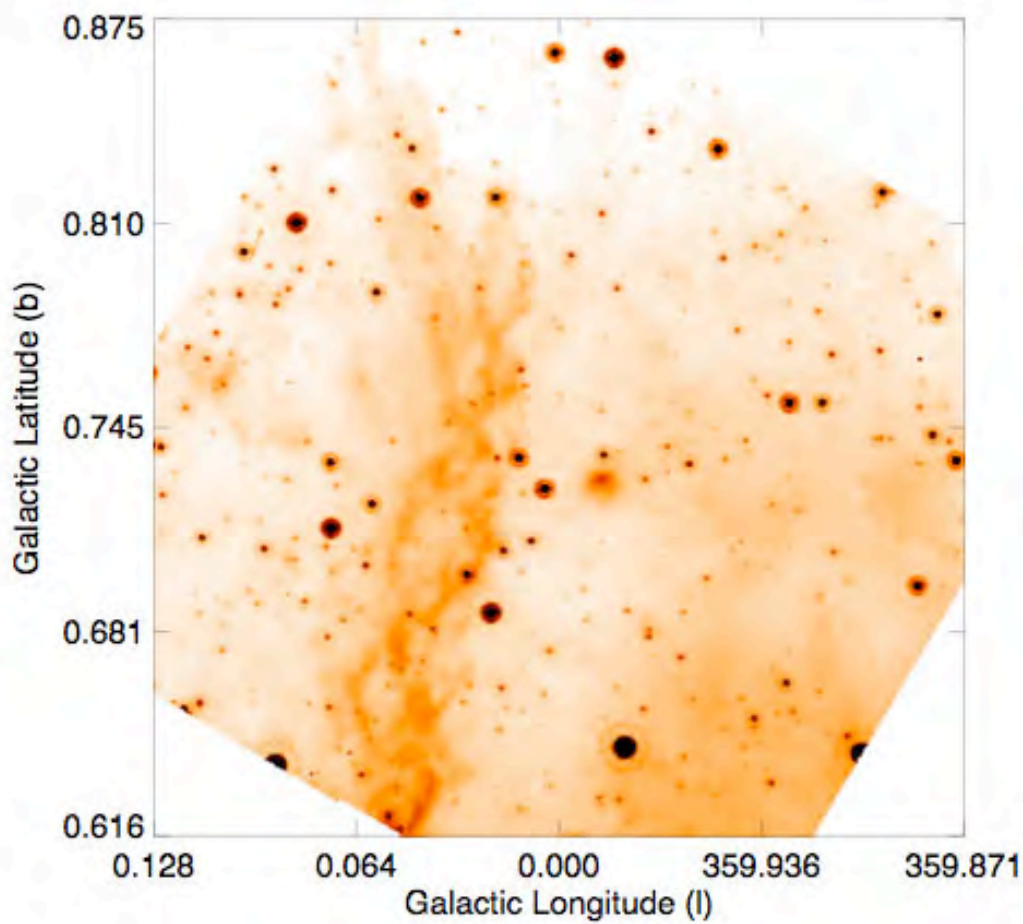

Figure 1:  The Double Helix Nebula, observed at the infrared wavelength of 24 μm with the MIPS camera on the Spitzer Space Telescope.  The spatial resolution is 6 arcseconds.  At the 8 kiloparsec distance of the Galactic center, 1 arcminute corresponds to 2.5 parsecs.  The full region observed extends well to the lower right of the region shown, and consists of a long strip centered on the bright infrared source AFGL5376[7], upon which we will report separately.

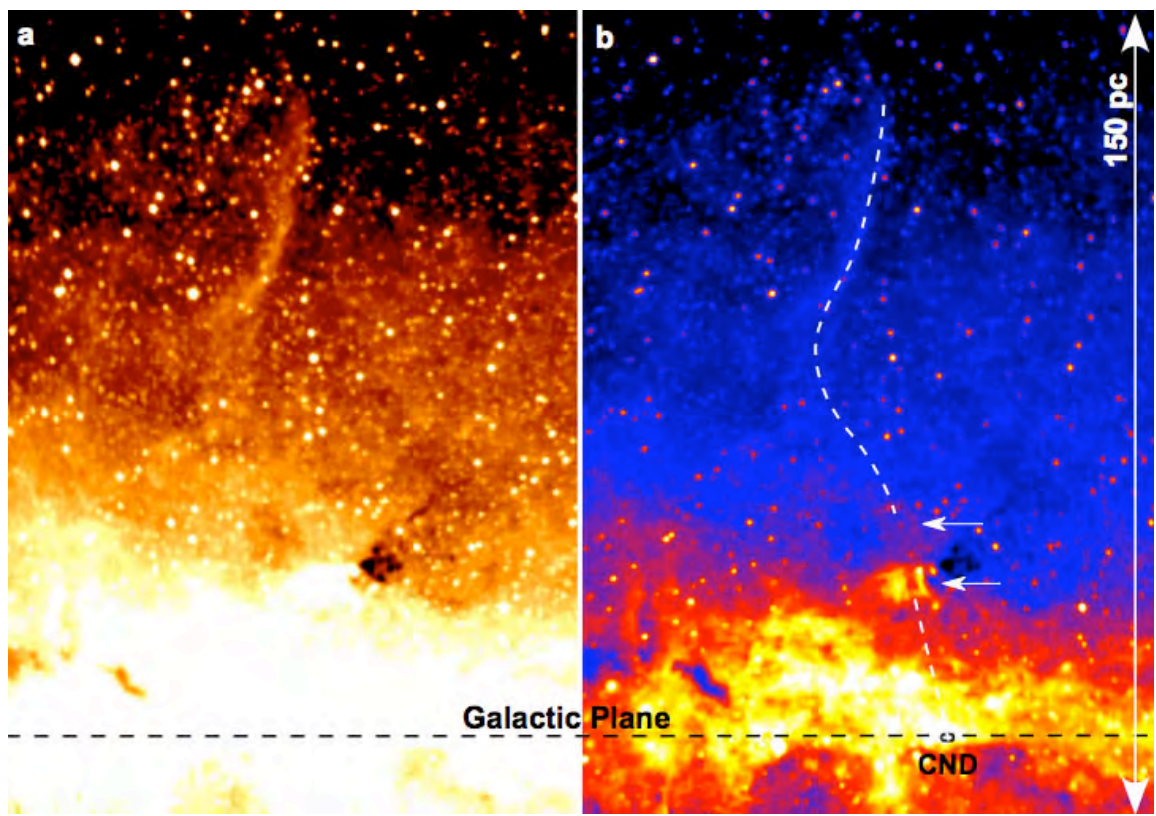

Figure 2: Images of a 1° field (from -0.1 to +0.9 degrees in Galactic latitude) that include the Galactic plane, the Galactic center, and the Double Helix Nebula. These data were taken with the Midcourse Space Experiment (MSX) satellite. The spatial resolution of these images is 20 arcseconds. Left: A-band image (8.3 μm) with a color scale chosen to emphasize the DHN. Right: same image, with a different color scale. The relative locations and sizes of the DHN, the circumnuclear disk (CND), and the proposed channel linking them, are all shown, along with the two bright channel segments marked with arrows and described in the text. These channel segments are located at Galactic coordinates (degrees) $l$, $b$=-0.01, 0.15 and 0.0, 0.23. Additional MSX images of the channel segments are available in the supplementary online material.

***Supplementary Information*** *(for on-line posting by Nature)*

*Additional Discussion*

<u>Alternatives to thermal dust emission to account for the spectral energy distribution:</u>

**Thermal bremsstrahlung emission from hot electrons.**  Assuming a power-law spectrum in the infrared with surface brightness proportional to $\nu^\gamma$, where $\nu$ is the frequency, we find that the best-fit value for $\gamma$ falls consistently in the range $+0.78 \pm 0.15$. For bremsstrahlung emission, even in the most favorable, optically thin regime, for which an electron density of 2000 cm$^{-3}$ and a total ionized mass of 50 M$_\odot$ are required, the spectrum should be flat ($\gamma \sim$ -0.1), rather than rising strongly toward shorter wavelengths. If the optical depth were large enough to explain the spectral slope, the density would be unreasonably high ($\sim 5 \times 10^7$ cm$^{-3}$).

**Nonthermal emission from a population of relativistic electrons.**  Nonthermal models are unattractive because the spectral slope in the infrared would imply a steeply rising electron energy distribution, $n(E) \propto E^{0.6}$, for which there are few precedents.

We conclude that the these alternative mechanisms to account for the 8 to 24 µm emission from the Double Helix Nebula are unlikely, and that the preferred mechanism is thermal emission from very warm dust grains.  Additional surface brightness measurements with the IRAC camera on Spitzer will provide enough information to explore these alternatives in detail.

*Supplementary Material:* <u>Table 1:</u> MSX surface brightnesses[1] and parameter fits for selected nebular positions. These representative positions were chosen to lie on the strands of the Double Helix Nebula (DHN), as inferred from the Spitzer/MIPS data.

| | band A (8.28 μm) | | band C (12.13 μm) | | band D (14.65 μm) | | band E (21.34 μm) | | fitted parameters | | |
|---|---|---|---|---|---|---|---|---|---|---|---|
| L, B (deg.)[2] | $\Sigma_{SB}$ | DHN excess[3] | $\Sigma_{SB}$ | DHN excess | $\Sigma_{SB}$ | DHN excess | $\Sigma_{SB}$ | DHN excess | $T_d$ (K)[4] p=1 | $T_d$ (K) p=2 | $\alpha$[5] |
| 0.084 0.544 | 6.33 | 1.50 | 6.22 | 1.49 | 2.73 | 0.58 | 2.73 | 0.59 | 403 | 306 | 0.77 |
| 0.064 0.588 | 6.13 | 2.10 | 6.06 | 2.18 | 2.49 | 0.74 | 2.24 | 0.48 | 408 | 308 | 0.85 |
| 0.044 0.628 | 5.62 | 1.57 | 5.53 | 1.51 | 2.13 | 0.25 | 2.22 | 0.45 | 428 | 320 | 0.92 |
| 0.053 0.676 | 5.73 | 1.83 | 5.58 | 1.80 | 2.27 | 0.73 | 1.90 | 0.32 | 418 | 314 | 0.91 |
| 0.026 0.718 | 5.15 | 1.50 | 5.00 | 1.31 | 1.89 | 0.44 | 2.47 | 1.14 | 402 | 309 | 0.54 |
| 0.029 0.744 | 5.13 | 1.52 | 5.04 | 1.50 | 2.07 | 0.57 | 1.80 | 0.36 | 415 | 312 | 0.87 |
| 0.021 0.774 | 4.15 | 0.82 | 4.14 | 0.85 | 1.87 | 0.44 | 1.71 | 0.43 | 379 | 291 | 0.61 |

[1]Units for surface brightness, $S_{SB}$, (in-band radiance) are $10^{-6}$ W m$^{-2}$ ster$^{-1}$

[2]Galactic coordinates, in degrees

[3]The DHN excess is the value of $S_{SB}$ in excess of the local background

[4]$T_d$ is the dust temperature assuming that the dust emissivity is proportional to (frequency)$^p$

[5]$\alpha$ is the best-fit power-law spectral index, $\Sigma_{SB} \sim \nu^\alpha$.

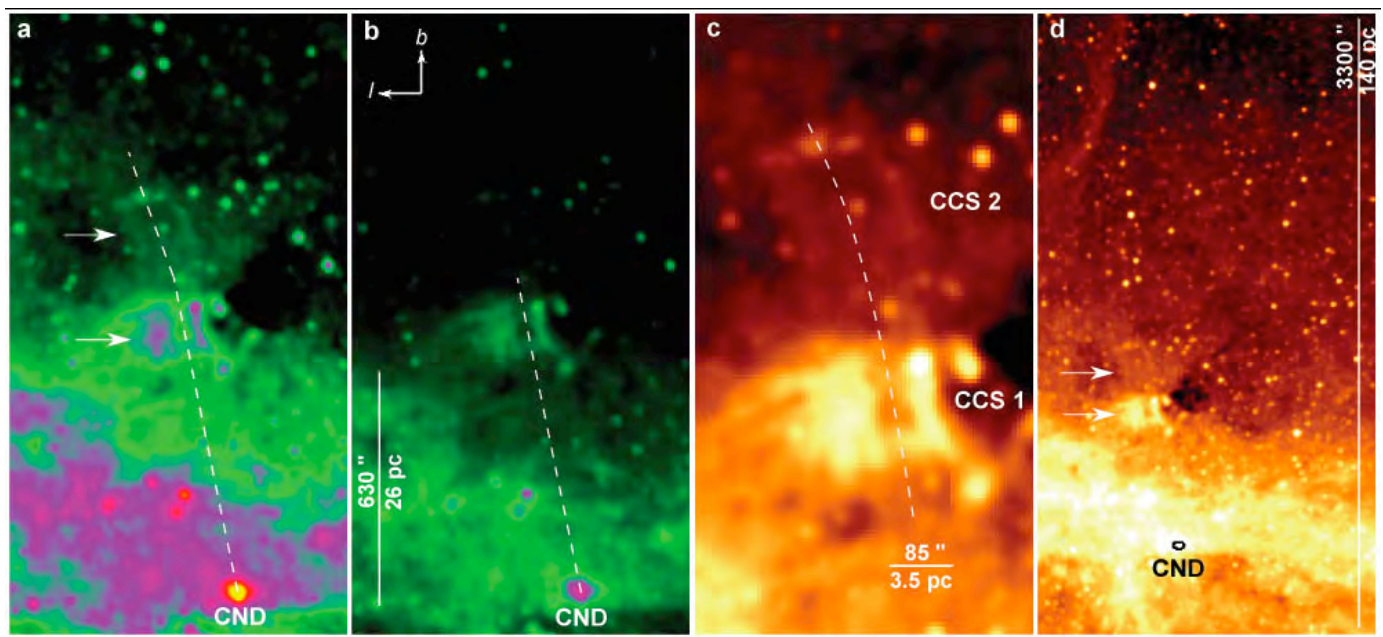

<u>Legend for Supplementary Figure:</u>  Four views of the cylindrical channel segments that lie along the hypothesized channel connecting the circumnuclear disk (CND) with the Double Helix Nebula (DHN).  All data shown are band C data (12.1 µm) from the MSX satellite database (http://irsa.ipac.caltech.edu/Missions/msx.html).  Horizontal arrows on panels *a* and *d* show the locations of these segments, and in the most magnified image, panel c, these are labeled as CCS1 and CCS2.  Panels *a* and *b* show the same field with different color stretches in order to illustrate the intensity contrast between CCS1 and CCS2.  Panel *d* presents the largest field, which includes the DHN at upper left.  The dashed lines in panels *a*, *b*, and *c* show the projected axis of the channel along which the torsional wave is hypothesized to propagate.  The angular and physical scales are shown on each figure with scale bars.  Note that the width of the cylindrical channel segments is very similar to that of both the DHN and the CND.